\documentclass[superscriptaddress,aps,preprintnumbers,amsmath,showpacs,amssymb,prd,nofootinbib,reprint]{revtex4-1}

\usepackage[dvipdfmx]{graphicx}
\usepackage{amsfonts}
\usepackage[mathscr]{eucal}
\usepackage{ascmac}
\usepackage{epstopdf}
\usepackage{dcolumn}
\usepackage{bm}
\usepackage[colorlinks=true,linkcolor=blue,citecolor=blue]{hyperref}
\usepackage{hyperref}
\usepackage{color}

\begin{document}


\newcommand{\vev}[1]{ \left\langle {#1} \right\rangle }
\newcommand{\bra}[1]{ \langle {#1} | }
\newcommand{\ket}[1]{ | {#1} \rangle }
\newcommand{\EV}{ \ {\rm eV} }
\newcommand{\KEV}{ \ {\rm keV} }
\newcommand{\MEV}{\  {\rm MeV} }
\newcommand{\GEV}{\  {\rm GeV} }
\newcommand{\TEV}{\  {\rm TeV} }
\newcommand{\1}{\mbox{1}\hspace{-0.25em}\mbox{l}}
\newcommand{\Red}[1]{{\color{red} {#1}}}

\newcommand{\lmk}{\left(}  
\newcommand{\rmk}{\right)}
\newcommand{\lkk}{\left[}  
\newcommand{\rkk}{\right]}
\newcommand{\lhk}{\left \{ }  
\newcommand{\rhk}{\right \} }
\newcommand{\del}{\partial}  
\newcommand{\la}{\left\langle} 
\newcommand{\ra}{\right\rangle}
\newcommand{\half}{\frac{1}{2}}

\newcommand{\bea}{\begin{array}}
\newcommand{\eea}{\end{array}}
\newcommand{\beq}{\begin{eqnarray}}
\newcommand{\eeq}{\end{eqnarray}}

\newcommand{\dd}{\mathrm{d}}
\newcommand{\Mpl}{M_{\rm Pl}}
\newcommand{\mg}{m_{3/2}}
\newcommand{\abs}[1]{\left\vert {#1} \right\vert}
\newcommand{\mphi}{m_{\phi}}
\newcommand{\Hz}{\ {\rm Hz}}
\newcommand{\for}{\quad \text{for }}
\newcommand{\Min}{\text{Min}}
\newcommand{\Max}{\text{Max}}
\newcommand{\Kahler}{K\"{a}hler }
\newcommand{\cphi}{\varphi}
\newcommand{\Tr}{\text{Tr}}
\newcommand{\diag}{{\rm diag}}

\newcommand{\SUf}{SU(3)_{\rm f}}
\newcommand{\Upq}{U(1)_{\rm PQ}}
\newcommand{\Zpq}{Z^{\rm PQ}_3}
\newcommand{\Cpq}{C_{\rm PQ}}
\newcommand{\ubar}{u^c}
\newcommand{\dbar}{d^c}
\newcommand{\ebar}{e^c}
\newcommand{\nubar}{\nu^c}
\newcommand{\Ndw}{N_{\rm DW}}
\newcommand{\Fpq}{F_{\rm PQ}}
\newcommand{\fpq}{v_{\rm PQ}}
\newcommand{\Br}{{\rm Br}}
\newcommand{\Lag}{\mathcal{L}}
\newcommand{\Lqcd}{\Lambda_{\rm QCD}}
\newcommand{\cm}{{\rm \ cm}}

\newcommand{\qel}{\hat{q}_{\rm el}}
\newcommand{\ksplit}{k_{\text{split}}}
\def\GDM{\Gamma_{\text{DM}}}
\def\Gsplit{\Gamma_{\text{split}}}

\def\mg{m_{3/2}}
\def\Im{{\rm Im}}
\def\bea{\begin{array}}
\def\eea{\end{array}}
\def\Mpl{M_{\text{Pl}}}
\def\M{M_{\text{Pl}}}
\def\mN{m_{\text{NLSP}}}
\def\Td{T_{\text{decay}}}
\def\mphi{m_{\phi}}
\def\tanb{\text{tan}\beta}
\def\signmu{\text{sign}[\mu]}
\def\fb{\text{ fb}}
\def\ij{_{ij}}
\def\k{\lmk {\bf k} \rmk}
\def\tk{\lmk \tau, {\bf k} \rmk}
\def\xk{\lmk x, {\bf k} \rmk}
\def\TT{T_{ij}^{\rm TT}}
\def\Hz{\ {\rm Hz}}
\def\for{\quad \text{for }}
\def\Min{\text{Min}}
\def\Max{\text{Max}}
\def\Kahler{K\"{a}hler }
\def\cphi{\varphi}
\def\Log{\text{Log}}



\preprint{
IPMU 15-0189; DESY 15-195
}

\title{
Affleck-Dine leptogenesis and its backreaction to inflaton dynamics
}

\author{
Masaki Yamada
}
\affiliation{
ICRR, 
The University of Tokyo, 
Kashiwa, Chiba 277-8582, Japan}
\affiliation{Kavli IPMU (WPI), UTIAS, 
The University of Tokyo, 
Kashiwa, Chiba 277-8583, Japan}
\affiliation{
Deutsches Elektronen-Synchrotron DESY, 
22607 Hamburg, Germany
}

\date{\today}

\begin{abstract}  
We investigate the backreaction of the Affleck-Dine leptogenesis to inflaton dynamics 
in the $F$-term hybrid and chaotic inflation models in supergravity. 
We determine the lightest neutrino mass in both models 
so that the predictions of spectral index, tensor-to-scalar ratio, 
and baryon abundance are consistent with observations. 
\end{abstract}


\maketitle

\section{\label{introduction}Introduction}

The success of the Big Bang nucleosynthesis theory requires 
a large amount of baryon asymmetry 
at least at the temperature of $1 \MEV$ in the early Universe. 
However, 
the baryon asymmetry must be washed out due to the primordial inflation, 
so that 
there has to be some mechanism to generate the baryon asymmetry after inflation. 
The Affleck-Dine baryo/leptogenesis is a promising candidate of baryogenesis 
in supersymmetric (SUSY) theories~\cite{AD, Murayama:1993em, DRT}. 
A $B-L$ charged scalar field with a flat potential, called an Affleck-Dine (AD) field, 
can obtain a large tachyonic effective mass 
and have a large vacuum expectation value (VEV) during and after inflation. 
As the energy of the Universe decreases, 
the effective mass becomes inefficient 
and the AD field starts to oscillate coherently around the origin of its potential. 
At the same time, 
the phase direction of the AD field is kicked by its A-term potential. 
Since the $B-L$ number density is proportional to the phase velocity of the AD field, 
the $B-L$ asymmetry is generated through this dynamics. 
Finally, the coherent oscillation decays and dissipates into thermal plasma 
and then the $B-L$ asymmetry is converted to the desired baryon asymmetry 
through the sphaleron effects~\cite{Kuzmin:1985mm, Fukugita:1986hr}.

Since the AD field obtains a large VEV during inflation, 
we should take into account its effect on inflaton dynamics via supergravity effects. 
In fact, 
there are many works revealing that 
a constant term in superpotential 
and a scalar field with a large VEV 
may affect inflaton dynamics~\cite{Lin:2006xta, 
Buchmuller:2014epa, Buchmuller:2014vda, Harigaya:2015pea}. 
These effects may rescue 
the $F$-term hybrid and chaotic inflation models, 
which themselves are somewhat inconsistent with the observations of CMB temperature fluctuations. 
In this letter, 
we apply their calculation to the scenario of the Affleck-Dine leptogenesis, 
focusing on the $L H_u$ flat direction in the minimal SUSY standard model sector. 
We show that 
the backreaction of the AD field on the inflaton dynamics 
can 
rescue the $F$-term hybrid and chaotic inflation models 
and the baryon asymmetry can be consistent with the observation at the same time. 
We predict extremely small mass for the lightest neutrino, 
which allows us to calculate the effective Majorana mass for the $0 \nu \beta \beta$ decay process.

\section{\label{ADBG}Affleck-Dine leptogenesis}

\subsection{\label{potential}Potential of the AD field}

Let us focus on the dynamics of the $L H_u$ flat direction: $\phi^2 = (L_i H_u) / \sqrt{2}$, 
where $L_i$ and $H_u$ are left-handed slepton with a family index $i$ and up-type Higgs, respectively. 
Since the observations of neutrino oscillation imply 
nonzero masses of neutrinos, 
we introduce a superpotential of 
\beq
 W^{(\rm AD)} &=& \frac{m_{\nu_i}}{2 \la H_u \ra^2} \lmk L_i H_u \rmk^2, \\ 
 &\equiv& \frac{\lambda}{4 \Mpl} \phi^4, 
 \label{W_phi}
\eeq
where $\la H_u \ra = \sin \beta \times 174 \GEV$ and $\tan \beta \equiv \la H_u \ra / \la H_d \ra$. 
We take the mass basis where the mass matrix for the neutrinos is diagonal. 
Here, the flat direction corresponding to the lightest neutrino is most important for the purpose of the Affleck-Dine leptogenesis, 
so that we identify that direction as the AD field and denote it as $\phi$. 
The lightest left-handed neutrino mass is then given by 
\beq
 m_{\nu} 
 &\simeq& 5.1 \times 10^{-10} \EV \lmk \frac{\lambda}{8.2 \times 10^{-5} } \rmk. 
\eeq
The coupling constant $\lambda$ is determined to account for the 
observations of baryon asymmetry and CMB temperature fluctuations.

The relevant potential of the AD field $\phi$ is written as 
\beq
 V_\phi 
 &=& 
 V_F + V_{\rm soft}  + V_H + V_T, 
\eeq
where 
\beq 
 V_F = \abs{ \frac{\del W^{(\rm AD)} }{ \del \phi }}^2, 
 \label{V_F}
\eeq
is the $F$-term potential. 
The potential $V_{\rm soft}$ represents the Higgs $\mu$ term and soft SUSY breaking terms in low energy: 
\beq 
 V_{\rm soft} = m_\phi^2 \abs{\phi}^2 
 + \lmk a m_{3/2} \lambda \frac{\phi^4}{4 \Mpl} + {\rm c.c.}  \rmk, 
 \label{soft terms}
\eeq
where $m_\phi$ ($= \mathcal{O}(1) \TEV$) is the mass of the $L H_u$ flat direction 
and $m_{3/2}$ is gravitino mass. 
We expect that the coefficient of A-term $a$ is of order unity.

The potential of $V_H$ 
is a so-called Hubble-induced mass term, 
which comes from supergravity effects~\cite{DRT}. 
In supergravity, the potential of scalar fields is given by 
\beq
 V_{\rm SUGRA} 
 = e^{K/\Mpl^2} \lkk \lmk D_i W \rmk K^{i \bar{j}} \lmk D_j W \rmk^* - \frac{3}{\Mpl^2} \abs{W}^2 \rkk, 
 \nonumber\\
 \label{SUGRA potential} 
\eeq
where $K$ is a \Kahler potential 
and $D_i W \equiv W_i + K_i W / \Mpl^2$. 
The subscripts represent derivatives with respect to corresponding fields 
and $K^{i \bar{j}} \equiv (K_{i \bar{j}})^{-1}$. 
In order to realize the Affleck-Dine leptogenesis, 
we assume 
\beq
 K = \abs{\phi}^2 + \abs{S}^2 + c \frac{1}{\Mpl^2} \abs{\phi}^2 \abs{S}^2, 
 \label{Kahler}
\eeq
where $S$ is an inflaton superfield and $c$ is an $O(1)$ constant. 
When we consider $F$-term inflation models, 
where the $F$-term of inflaton satisfies $\abs{F_S}^2 \simeq 3 H^2_{\rm inf} \Mpl^2$ 
with $H_{\rm inf}$ being the Hubble parameter during inflation, 
we obtain a Hubble-induced mass of the AD field: 
\beq
 V_H &=& 
 c_H H^2 \abs{\phi}^2 
 \label{Hubble-induced mass}
 \\
 c_H &=& 3 (1-c). 
\eeq
To realize the Affleck-Dine leptogenesis, we assume $c_H < 0$.

After inflation ends, 
the inflaton gradually decays into radiation 
and a background plasma develops with a temperature of 
\beq
 T = \lmk \frac{36 H(t) \Gamma_I \Mpl^2}{g_* (T) \pi^2} \rmk^{1/4} 
 ~~~~{\rm for}~~ T \gtrsim T_{\rm RH}, 
 \label{T during osc.}
\eeq
where $g_* (T)$ is the effective number of relativistic degrees of freedom in the thermal plasma. 
The decay rate of inflaton $\Gamma_I$ is related with the reheating temperature $T_{\rm RH}$ as 
\beq
 T_{\rm RH} \simeq \lmk \frac{90}{g_* (T_{\rm RH}) \pi^2 } \rmk^{1/4} \sqrt{\Gamma_I \Mpl}. 
\eeq
The AD field acquires a thermal potential via 2-loop effect 
when its VEV is larger than the temperature: 
\beq
 V_T 
 = 
 c_T \alpha_s^2 T^4 \log \lmk \frac{\abs{\phi}^2}{T^2} \rmk, 
\eeq
where $c_T = 45/32$ 
and $\alpha_s \equiv g_s^2 / 4 \pi$ is the strong coupling constant~\cite{Anisimov:2000wx, Fujii:2001zr}.

\subsection{\label{dynamics}Dynamics of the AD field}

When we consider $F$-term inflation models, 
the Hubble induced mass term of Eq.~(\ref{Hubble-induced mass}) arises during inflation. 
Since we assume $c_H < 0$, 
the AD field stays at the following potential minimum: 
\beq
 \la \abs{\phi} \ra_{\rm inf} \simeq 
 \lmk \sqrt{\frac{\abs{c_H}}{3}} \frac{H_{\rm inf} \Mpl}{\lambda} \rmk^{1/2}. 
 \label{AD VEV}
\eeq

After inflation ends, 
its VEV is determined as Eq.~(\ref{AD VEV}) with 
the replacement of $H_{\rm inf} \to H(t)$ during the inflaton oscillation dominated era. 
Note that the phase direction of the AD field stays at a certain phase due to the Hubble friction effect. 
When the Hubble parameter decreases to $m_\phi$ or $(\phi^{-1} V_T')^{1/2}$, 
the AD field starts to oscillate around the origin of the potential. 
We denote the Hubble parameter as $H_{\rm osc}$: 
\beq
 H_{\rm osc} \simeq \Max \lkk m_\phi, \ \sqrt{\phi_{\rm osc}^{-1} V_T'} \rkk, 
 \label{H_osc}
\eeq
where $\phi_{\rm osc}$ is the VEV of the AD field at the beginning of oscillation. 
At the same time, the AD field starts to rotate in the complex plane 
because its phase direction is kicked by the A-term of Eq.~(\ref{soft terms}). 
This is the dynamics that generates $B-L$ asymmetry. 
The amplitude of the flat direction decreases as time evolves 
due to the Hubble friction effect 
and the $B-L$ breaking effect of Eq.~(\ref{soft terms}) becomes irrelevant soon after the beginning of oscillation. 
Thus, the generated $B-L$ number is conserved soon after the AD field is kicked in the 
complex plane. 
We numerically solve the equation of motion 
and find that the $B-L$ number density at the beginning of oscillation is given by 
\beq
 n_{B-L} (t_{\rm osc}) &\equiv& 
 \epsilon H_{\rm osc} \phi_{\rm osc}^2 
 \\
 \epsilon &=& (0.2-1.7) \times a \sin \lmk n \theta_0 \rmk \frac{m_{3/2}}{H_{\rm osc}} 
 \\
 &\equiv& \tilde{\epsilon} \frac{m_{3/2}}{H_{\rm osc}}, 
\eeq
where $\theta_0$ is an initial phase of the AD field. 
Here, we define the ellipticity parameter $\epsilon$ ($\le 1$), 
which 
represents the efficiency of the Affleck-Dine mechanism.

Finally, the coherently oscillating AD field decays and dissipates into thermal plasma~\cite{Mukaida:2012qn} 
and the sphaleron effect converts 
the $B-L$ asymmetry to baryon asymmetry~\cite{Kuzmin:1985mm, Fukugita:1986hr}. 
The resulting baryon-to-entropy ratio $Y_b$ is given by 
\beq
 Y_b  
&\equiv&
 \frac{n_b}{s} 
\simeq 
 \frac{8}{23} \frac{\epsilon T_{\rm RH}}{4 H_{osc}} \lmk \frac{\phi_{osc}}{\Mpl} \rmk^2, 
 \label{Y_b}
 \\
 &\simeq& 
 6.5 \times 10^{-11} \tilde{\epsilon} \lmk \frac{\lambda}{10^{-4}} \rmk^{-3/2} 
 \lmk \frac{m_{3/2}}{100 \GEV} \rmk, 
 \label{Y_b 2}
\eeq
where $8/23$ in the first line is the sphaleron factor~\cite{Harvey:1990qw}. 
In the second line, we assume $\alpha_s \sqrt{\lambda} T_{\rm RH} \gtrsim m_\phi$ 
to use $H_{\rm osc} \simeq \sqrt{\phi_{\rm osc}^{-1} V_T'}$ in Eq.~(\ref{H_osc}). 
Note that the result is independent of the reheating temperature~\cite{Fujii:2001zr}. 
The observed baryon asymmetry of 
$Y_b^{\rm obs} \simeq 8.6 \times 10^{-11}$~\cite{pdg} 
can be explained when the coupling $\lambda$ satisfies 
\beq
 \lambda \simeq 8.2 \times 10^{-5} \tilde{\epsilon}^{2/3} \lmk \frac{m_{3/2}}{100 \GEV} \rmk^{2/3}. 
\eeq

\section{\label{backreaction}Backreaction to inflaton dynamics}

\subsection{\label{$F$-term}$F$-term hybrid inflation}

In this subsection, we consider the simplest model of $F$-term hybrid inflation~\cite{Copeland:1994vg, Dvali:1994ms} 
taking into account the effect of the AD field on the dynamics of inflaton. 
The superpotential of the inflaton sector is given by 
\beq
 W^{(\rm inf)}
 = 
 \kappa S \lmk \psi \bar{\psi} - \frac{v^2}{2} \rmk, 
\eeq
where $\kappa$ is a coupling constant, 
$S$ is inflaton, and $\psi$ and $\bar{\psi}$ are waterfall fields. 
When the inflaton $S$ has a sufficiently large VEV, 
the waterfall fields obtain large effective masses of $\kappa \la S \ra$ 
and thus stay at the origin of the potential. 
The inflaton $S$ obtains the Coleman-Weinberg potential of 
\beq
 V_{\rm CW}
 \simeq 
 \frac{\kappa^4 v^4}{32 \pi^2} \ln \lmk \frac{\abs{S}}{S_{\rm cr}} \rmk, 
 \label{CW}
\eeq
where $S_{\rm cr} \equiv v / \sqrt{2}$. 
The inflaton $S$ slowly rolls down to the origin of the potential until its VEV reaches the critical value of $S_{\rm cr}$. 
During this slow roll, 
the energy density is dominated by the $F$-term potential energy of $\kappa^2 v^4/4$, 
so that inflation occurs. 
When the inflaton $S$ reaches a critical VEV of $S_{\rm cr}$, 
the waterfall fields and inflaton start to oscillate about their global minimum 
and inflation ends. 
In this simplest model, 
the spectral index is predicted as $n_s \simeq 1- 1/ \mathcal{N}_* \simeq 0.98$, 
where $\mathcal{N}_*$ ($\approx 55$) is the e-folding number at the horizon exit of the CMB scale. 
This prediction is inconsistent with the observed value more than $2$ sigma level: 
$n_s^{(\rm obs)} = 0.963 \pm 0.008$~\cite{Ade:2013xla}.

Now we take into account the backreaction of the AD field 
to the dynamics of the inflaton. 
In supergravity, the potential of scalar fields is determined by Eq.~(\ref{SUGRA potential}). 
When we consider the total superpotential $W = W^{(\rm AD)} + W^{(\rm inf)}$, 
the terms of $W_S K^{\bar{S} \phi} W_{\bar{\phi}} + {\rm c.c.} - 3 \abs{W}^2$ 
give a linear potential of inflaton such as~\cite{Buchmuller:2014epa}
\beq
 V_{\rm BR} 
 \simeq 
 a' \frac{\kappa v^2 }{\Mpl^2} 
 \la W^{(\rm AD)} \ra S + {\rm c.c.}, 
\eeq
where $a'$ is an $O(1)$ constant determined by higher dimensional Kahler potentials 
and 
$\la W^{(\rm AD)} \ra$ is determined by Eqs.~(\ref{W_phi}) and (\ref{AD VEV}). 
Hereafter we assume $a'=1$.

The effect of the linear term in the $F$-term hybrid inflation model has been studied 
in Ref.~\cite{Buchmuller:2014epa}. 
They have found that 
the linear term affect the inflaton dynamics 
when the slope of the linear term is the same order with that of the Coleman-Weinberg potential. 
They introduce a parameter to describe the relative importance of the two contributions to the slope: 
\beq
 \xi \equiv \frac{2^{9/2} \pi^2}{\kappa^3 \ln 2} \frac{\la W^{(\rm AD)} \ra}{v \Mpl^2}, 
\eeq
which should be smaller than unity so that the inflaton can rolls towards the critical value. 
When $\xi$ is of order but below unity, 
the linear term is efficient for the inflaton dynamics. 
We define 
a critical value of coupling constant for the AD field such as 
\beq
 \lambda_c \equiv 2.2 \frac{v^3}{\kappa}, 
 \label{lambda_c}
\eeq
where we use $H_{\rm inf}^2 = \kappa^2 v^4 / 12 \Mpl^2$. 
When $\lambda$ is near the critical value, 
$\xi$ is close to unity 
and the backreaction of the AD field to inflaton dynamics is efficient. 
Note that 
$\lambda$ should not larger than $\lambda_c$ 
so that the inflaton can rolls towards the critical value and inflation can ends.

\begin{figure}[t]
\centering 
\begin{tabular}{l l}
\includegraphics[width=.40\textwidth, bb=0 0 360 341]{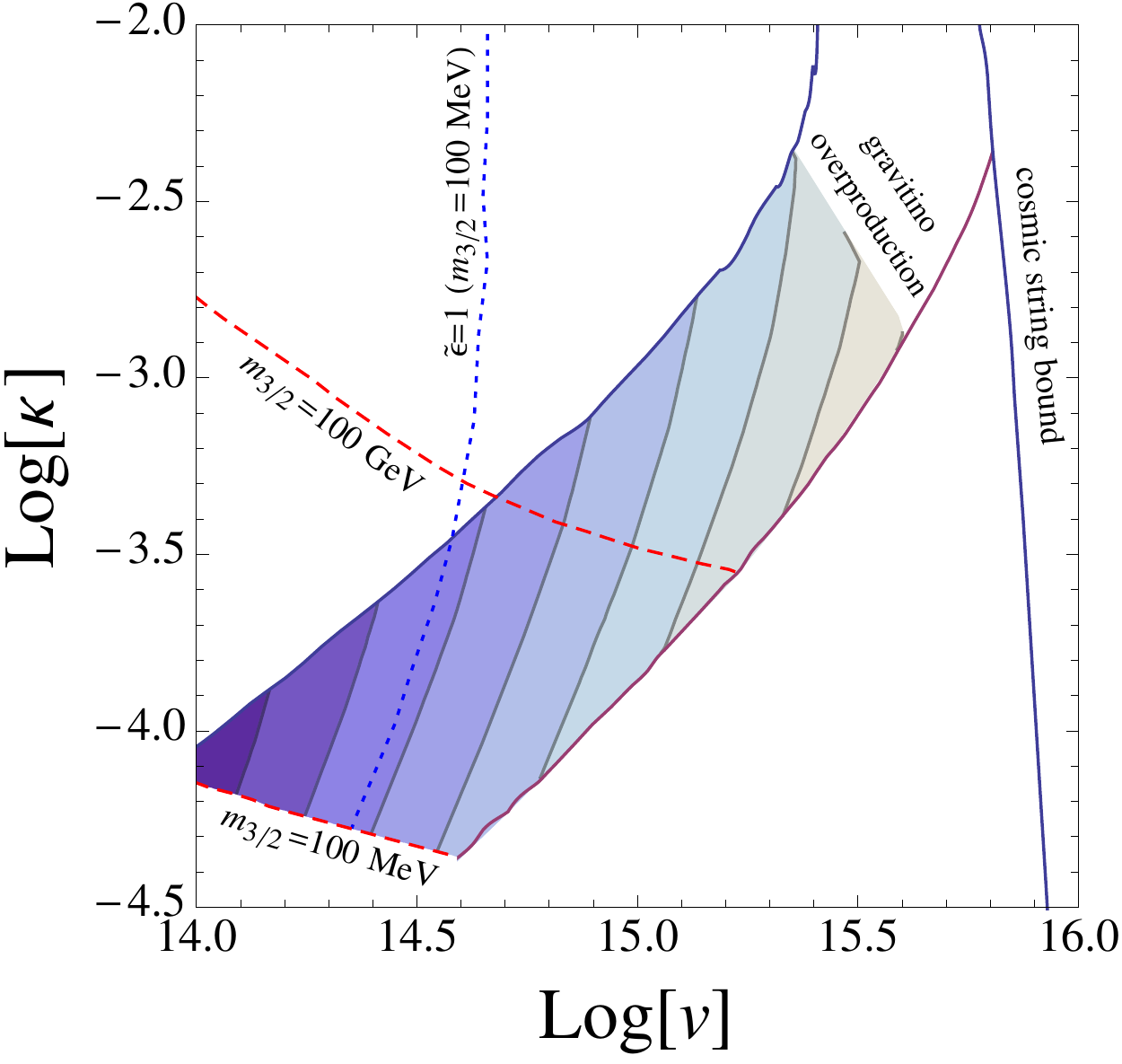} 
\includegraphics[width=.05\textwidth, bb=0 -40 39 225]{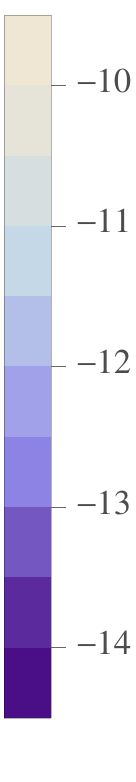} 
\end{tabular}
\caption{
Contour plot of neutrino mass in the unit of $\EV$ in $\Log[v]$-$\Log [\kappa]$ plane. 
For the case of $m_{3/2} = 100 \GEV$, 
the spectral index as well as the baryon abundance can be consistent with the observations 
above the corresponding red-dashed curve in the colored region, 
while 
for the case of $m_{3/2} = 100 \MEV$, 
they can 
above the corresponding red-dashed curve and blue-dotted curve in the colored region. 
}
  \label{fig1}
\end{figure}

Since the linear term breaks R-symmetry, 
under which the inflaton $S$ is charged, 
we need to investigate the inflaton dynamics in its complex plane 
as done in Ref.~\cite{Buchmuller:2014epa}.%
\footnote{
A CP-odd component of inflaton is excited via this dynamics, 
which also provide another scenario of baryogenesis~\cite{Takahashi:2015ula}. 
}
We read their result of Fig.~9, 
where 
desired values of $\la W^{(\rm AD)} \ra$ can be read from 
the contours of gravitino masses by the relation of $m_{3/2} \Mpl^2 \leftrightarrow \la W^{(\rm AD)} \ra$.%
\footnote{
The dynamics of the phase direction of the AD field can be neglected 
for the case of $\lambda \ll \kappa$~\cite{future work}, which is actually satisfied in our case, 
so that the dynamics of inflaton is basically equivalent to the one in Ref.~\cite{Buchmuller:2014epa}. 
}
The result is shown in Fig.~\ref{fig1}, 
where the spectral index as well as the baryon asymmetry 
can be consistent with the observed values in the colored region. 
Here, 
we assume that 
the final phase of the inflaton is larger than $\pi/32$ to avoid a fine-tuning of initial condition. 
Above the red-dashed curves for each case of gravitino mass, 
we can neglect the effect of a linear term arising from low energy SUSY breaking, 
which is investigated in the original work of Ref.~\cite{Buchmuller:2014epa}. 
If there is only the effect of a linear term arising from low energy SUSY breaking, 
the spectral index can be consistent with the observation 
on the red-dashed curve for each case of gravitino mass. 
Thus we can explain the observation on and above the red-dashed curve for each case of gravitino mass 
in our model.%
\footnote{
We neglect an $\mathcal{O}(1)$ uncertainty arising near the red-dashed curve 
that comes from the phase difference 
between two linear terms. 
}
The right region is excluded by the cosmic string bound~\cite{Ade:2013xla}, 
while 
the upper-right region is excluded by the overproduction of gravitinos 
via inflaton decay~\cite{Nakamura:2006uc} 
and/or thermal production~\cite{Bolz:2000fu, Pradler:2006qh}.

Since the value of superpotential of the AD field is determined 
at each point in Fig.~\ref{fig1}, 
we can determine its coupling constant $\lambda$. 
Then we can use Eq.~(\ref{Y_b 2}) to calculate the baryon abundance. 
For the case of $m_{3/2} = 100 \GEV$, 
we can explain the baryon abundance by taking $\tilde{\epsilon}$ properly. 
On the other hand, 
for the case of $m_{3/2} = 100 \MEV$, 
the baryon asymmetry cannot be produced efficiently 
below the blue-doted curve even if $\tilde{\epsilon}$ is as large as unity.%
\footnote{
When the coefficient of A-term $a$ in Eq.~(\ref{soft terms}) 
is much larger than unity, 
$\tilde{\epsilon}$ can be larger than unity 
and the bound of the blue curve disappear. 
}
We predict lightest neutrino mass $m_\nu$ as given in the contour plot. 
Since the coupling constant in the superpotential of the AD field 
is roughly determined by Eq.~(\ref{lambda_c}) 
to affect the inflaton dynamics, 
$m_\nu$ is larger for larger $v$ and smaller $\kappa$. 
From the figure, 
we can see that 
$m_\nu$ can be as large as $10^{-10} \EV$ for the case of $m_{3/2} = 100 \GEV$, 
while it is at most $10^{-13} \EV$ for the case of $m_{3/2} = 100 \MEV$.

\subsection{\label{chaotic}Chaotic inflation}

In this subsection, we consider a chaotic inflation model in supergravity 
where an inflaton superfield $I$ has $Z_2$ and shift symmetries in the Kahler potential~\cite{Kawasaki:2000yn}: 
\beq
 K^{(\rm inf)} = \frac{1}{2} \lmk I + I^* \rmk^2. 
\eeq
The imaginary part of its scalar component $\eta \equiv (I - I^*)/\sqrt{2}$ is identified with inflaton. 
The shift symmetry is explicitly broken by a superpotential of 
\beq
 W^{(\rm inf)} = m I S, 
\eeq
where $S$ is a stabiliser field with a Kahler potential of Eq.~(\ref{Kahler}). 
When the inflaton has a large VEV, 
the stabiliser field obtains a large effective mass 
and stays at the origin. 
The inflaton potential is then given by the quadratic potential via the $F$-term of the stabiliser field. 
Thanks to the shift symmetry in the Kahler potential, 
the VEV of inflaton can be larger than the Planck scale 
and quadratic chaotic inflation can be realized in this model. 

Here, we take into account the backreaction of the AD field. 
The full supergravity potential is given by 
\beq
 V &=& e^{\abs{\phi}^2/ \Mpl^2} 
 \lkk \frac{1}{2} m^2 \eta^2 \frac{1}{1+ c \abs{\phi}^2/ \Mpl^2} 
 \right.
 \nonumber
 \\
 &&\left.
 ~~~~~~ 
 +  \lambda^2 \lmk \frac{\abs{\phi}^{6}}{\Mpl^2} + \frac{5}{16} \frac{\abs{\phi}^8}{\Mpl^4} + \frac{1}{16} \frac{\abs{\phi}^{10}}{\Mpl^6} \rmk \rkk, 
 \label{V in chaotic}
\eeq
where $c$ is the parameter in the Kahler potential [see Eq.~(\ref{Kahler})]. 
This potential implies that 
the effect of the AD field is relevant 
when its VEV is as large as the Planck scale. 
Since $H_{\rm inf} \sim 10 m$ in the chaotic inflation model, 
the VEV of the AD field is as large as the Planck scale for the case of 
\beq
 \lambda \sim \lambda_c \equiv 10  \sqrt{c-1} \frac{m}{ \Mpl}, 
\eeq
[see Eq.~(\ref{AD VEV})].

\begin{figure}[t]
\centering 
\includegraphics[width=.40\textwidth, bb=0 0 360 344]{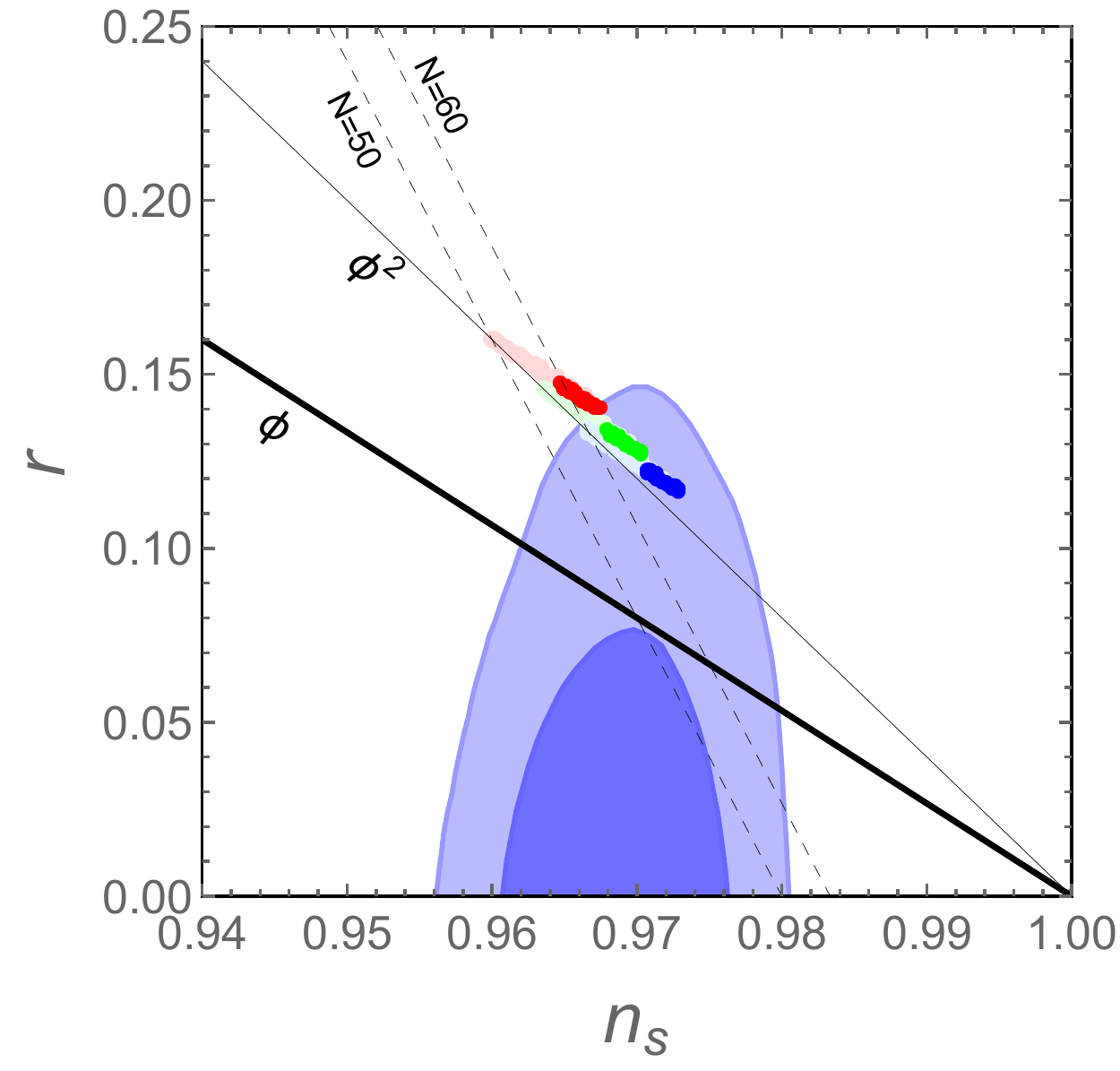} 
\caption{
Spectral index and tensor-to-scalar ratio in the chaotic inflation model with the backreaction of the AD field. 
The red, green, and blue dots represent our results at e-folding numbers of $50$, $55$, and $60$, respectively. 
We randomly take $100$ points for the parameters $c$ and $\lambda$ 
within the intervals of $[1,10]$ and $[0, 100 m/\Mpl]$, respectively. 
We plot the results as the light dots for the case of 
$\lambda / \lambda_c < 0.5 $, $5 < \lambda / \lambda_c$, or $c< 5$. 
The blue regions are 
the $1 \sigma$ (deep colored regions) and $2 \sigma$ (pale colored regions) constraints of the Planck experiment~\cite{Ade:2015xua}. 
For comparison with standard results, 
we plot the predictions in the chaotic inflation models with linear and quadratic potentials 
without the backreaction as the black thin and thick lines, respectively, 
where the results are given as intersection points of black lines and dashed lines 
for corresponding e-folding numbers. 
}
  \label{fig2}
\end{figure}

We numerically solve the equations of motion of the inflaton $\eta$ and the AD field $\phi$ 
and calculate the tensor-to-scalar ratio and spectral index. 
We show the result in Fig.~\ref{fig2}, 
where 
we take the parameters $c$ and $\lambda$ 
randomly within the intervals of $[1,10]$ and $[0, 100 m/\Mpl]$, respectively. 
The red, green, and blue dots represent 
the results at e-folding numbers of $50$, $55$, and $60$, respectively. 
As a result, 
the tensor-to-scalar ratio can be as small as $0.14$, $0.13$, and $0.12$ 
at the e-folding number of $50$, $55$, and $60$, respectively, 
which is marginally consistent with the present upper bound within $2 \sigma$. 
We plot the results as the light dots for the case of 
$\lambda / \lambda_c < 0.5 $, $5 < \lambda / \lambda_c$, or $c< 5$, 
which clarifies that 
the tensor-to-scalar ratio can be smaller only for the case of $0.5 < \lambda / \lambda_c < 5$ 
and $c>5$. 
This requires that the coupling constant in the superpotential is of order $10 m/\Mpl \sim 10^{-4}$, 
so that 
the lightest neutrino mass is predicted to be of order $10^{-9} \EV$. 
Note that the resulting baryon asymmetry of Eq.~(\ref{Y_b 2}) is naturally consistent with 
the observation when gravitino mass is of order $100 \GEV-1 \TEV$.

Finally, we also perform numerical calculations 
including higher dimensional Kahler potentials of 
\beq
 K \supset d \frac{1}{\Mpl^2} \abs{\phi}^4 + d' \frac{1}{\Mpl^4} \abs{\phi}^6 
 + c'  \frac{1}{\Mpl^4} \abs{S}^2 \abs{\phi}^4, 
\eeq
and find that the tensor-to-scalar ratio can not be smaller than about $0.11$ at the e-folding number of $60$ 
even in this case.%
\footnote{
We also take into account kinetic couplings between the inflaton and AD field 
due to the higher-dimensional \Kahler potential of $c'' \abs{\phi}^2 ( I + I^*)^2 / 2 \Mpl^2$. 
However, we find that their effect is also very limited. 
}
This is in contrast with the result of Ref.~\cite{Harigaya:2015pea}, 
where they have investigated the effect of an additional scalar field to chaotic inflation in a non-SUSY model 
and found that the tensor-to-scalar ratio can be much smaller than $0.1$. 
This is because 
the exponential factor in the supergravity potential of Eq.~(\ref{V in chaotic}) 
makes the VEV of the AD field smaller 
and its backreaction to the inflaton dynamics smaller in supergravity.

\section{\label{conclusion}Discussion and conclusions}

We have investigated the backreaction of the AD field to inflaton dynamics, 
focusing on the $L H_u$ flat direction in the minimal SUSY standard model. 
In the $F$-term hybrid inflation model, 
a linear term of inflaton potential is induced by the nonzero superpotential of the AD field. 
As a result, the spectral index as well as baryon abundance can be consistent with the observed values. 
In the chaotic inflation model with a shift symmetry in the inflaton Kahler potential, 
we have found that 
the tensor-to-scalar ratio can be as small as $0.12$ due to the backreaction of the AD field.

All of the above scenarios require 
a large VEV of the AD field during inflation. 
This is also favoured in light of avoiding the baryonic isocurvature constraint, 
which is particularly important in the chaotic inflation model~\cite{Enqvist:1998pf, Kasuya:2008xp, Harigaya:2014tla, Kawasaki:2015cla}. 
To realize a large VEV during inflation for the $L H_u$ flat direction, 
the mass of the lightest neutrino has to be extremely small. 
Thus 
the total neutrino mass is given by 
\beq
 \sum m_\nu \simeq 
 \left\{
 \bea{ll}
 0.06 \EV
 &~~~~\text{for \ NH} 
 \\
 0.1 \EV 
 &~~~~\text{for \ IH}, 
 \eea
 \right.
\eeq
for the cases of normal hierarchy (NH) and inverted hierarchy (IH), respectively. 
We can also 
calculate the upper and lower bounds on the effective Majorana mass 
for the $0 \nu \beta \beta$ decay process 
such as~\cite{Fujii:2001zr, Fujii:2002hp}
\beq
 0.001 \EV \lesssim \abs{m_{\beta \beta}} \lesssim 0.004 \EV 
 &&~~~~\text{for \ NH} 
 \\
 0.01 \EV \lesssim \abs{m_{\beta \beta}} \lesssim 0.04 \EV 
 &&~~~~\text{for \ IH}, 
\eeq
where we 
take the values for the experimentally measured parameters from Ref.~\cite{Capozzi:2013csa}. 
These results of total neutrino mass and effective Majorana mass 
are too small to measure in the near future at least for the case of NH. 
Therefore, if we would measure the total neutrino mass or 
the effective Majorana mass in the near future, 
we can falsify our scenario of the AD leptogenesis. 
On the other hand, 
if we would experimentally obtain only their upper bound 
and if 
the tensor-to-scalar ratio would be measured as $0.12-0.15$, 
our scenario of the AD leptogenesis after the chaotic inflation 
would be more attractive. 

Finally, let us comment on other scenarios of Affleck-Dine baryogenesis 
using other flat directions, such as the $u^c d^c d^c$ flat direction, 
where $u^c$ and $d^c$ are up-type and down-type right-handed squarks, respectively. 
In this case, 
there are some corrections in calculations of baryon abundance. 
The most important difference from our scenario 
is the possibility of the formation of non-topological solitons called Q-balls~\cite{
Coleman, Qsusy, EnMc, KuSh, KK}. 
In particular, 
as we have shown in this letter, 
the backreaction of the AD field is relevant when its VEV is sufficiently large during inflation, 
which implies that 
large Q-balls may form after the AD baryogenesis. 
In this case, 
Q-balls may decay after dark matter (the lightest SUSY particle) freezes out, 
so that their decay can be a non-thermal source of dark matter. 
There are interesting scenarios that 
the non-thermal production of dark matter from Q-ball decay 
can naturally explain 
the coincidence between the energy density of baryon and dark matter~\cite{Fujii:2001xp, 
Roszkowski:2006kw, ShKu, Kamada:2012bk} 
(see, e.g., Ref.~\cite{Harigaya:2014tla} 
in the case of chaotic inflation).

\vspace{1cm}

%
\section*{Acknowledgments}
M.Y. thanks W. Buchm\"{u}ller for grateful discussions and comments and for kind hospitality at DESY. 
This work is supported by World Premier International Research Center Initiative
(WPI Initiative), MEXT, Japan, 
the Program for the Leading Graduate Schools, MEXT, Japan, 
and the JSPS Research Fellowships for Young Scientists, No. 25.8715.
%

\vspace{1cm}



\end{document}